# THE GREAT CIRCLE CAMERA: A NEW DRIFT SCANNING INSTRUMENT

Dennis Zaritsky

UCO/Lick Observatory and Board of Astronomy and Astrophysics, Univ. of California, Santa Cruz, CA, 95064, Email: dennis@lick.ucsc.edu

Stephen A. Shectman and Gregory Bredthauer

Carnegie Observatories, 813 Santa Barbara St., Pasadena, CA, 91101, Email: shec@ociw.edu and greg@ociw.edu

## ABSTRACT

We discuss the design, construction, and use of a new class of scanning camera that eliminates a critical limitation of standard CCD drift-scan observations. A standard scan, which involves no correction for the differential drift rates and curved stellar paths across the field-of-view, suffers from severe image degradation even when one observes at moderate declinations. Not only does this effect limit the area of the sky over which drift scanning is viable, but as detector sizes increase, CCD mosaics become standard, and dome/telescope seeing improves, the area of sky for which scanning is acceptable (image degradation $\lesssim$ seeing) will be further reduced unless some action is taken. By modifying the scan path (the path on the sky traced by signal accumulated along a single CCD column) to lie along a great circle on the sky rather than along a path of constant declination, image degradation is minimized. In this paper, we discuss the design and implementation of a stage that rotates and translates the CCD during a drift-scan exposure so that the scan path is along a great circle on the sky. Data obtained during the commissioning run of the Great Circle Camera at the Las Campanas 1-m telescope are presented.

## 1. INTRODUCTION

Surveys of large areas of the sky provide unique information. However, the currently preferred optical detector, the charge-coupled device or CCD, is limited in physical size to a few centimeters and in resolution to about 2000 elements. These limits correspond to a field size of only about 30 arcmin at an angular resolution of 1 arcsec. Because of the relatively small field-of-view, large-area CCD surveys must consist of image mosaics, and so are inefficient. Nevertheless, the numerous advantages provided by CCDs (uniformity, linearity, and high quantum efficiency) justify the pursuit of techniques with which to improve the efficiency of CCD surveys. One such technique is referred to as drift scanning (cf. McGraw, Angel, & Sargent 1980).

At its simplest, the concept of drift scanning involves parking the telescope and observing the sky with a CCD while the Earth's motion moves celestial objects across the field-of-view of the telescope. To maintain the one-to-one correspondence of sky and image, the charge on the CCD which corresponds to a fixed point on the sky is moved across the CCD at precisely the same rate at which the sky is moving across the field-of-view. This drift rate is 15 seconds of arc per second at the celestial equator. The key efficiency advantage of drift scanning is that it minimizes the overhead associated with telescope positioning and CCD readout. A simple hypothetical observation demonstrates this advantage. Complete spatial coverage of a strip of sky that is 3° long and 23 arcmin wide at the celestial equator with a $2048^2$ CCD with 0.7 arcsec pixels (parameters chosen to match the observations that we describe below) will require eight pointed observations. Conservatively allowing one minute to offset the telescope and two minutes to read the CCD at each position, the overhead associated with these observations is 24 minutes. For a drift scan of the same region the overhead is the sum of the time of the initial offset (one minute) and the time required for the scan to "ramp up" (the scan only officially begins once the first object has crossed the entire field, which takes 96 sec at the celestial equator for this configuration). Therefore, the overhead associated with the scan is only about 2.5 minutes. In a large survey that requires many such strips, the overhead associated with pointed observations can easily add up to the equivalent of entire nights! The key scientific advantage of drift scanning is superior flat-fielding and image uniformity. The effect of pixel-to-pixel variations are minimized in a scan image because the signal is averaged over an entire column of pixels.

Drift scanning is not appropriate for every type of imaging project. A key disadvantage of scanning is that the exposure time is fixed to be the time it takes an object to cross the field-of-view. For a $2048^2$ CCD with 0.7 arcsec pixels that time is 96 sec at the celestial equator. One can rescan the same area of the sky to increase the total exposure time; however, the efficiency gain is then mitigated. Nevertheless, for relative shallow large-area surveys, drift scanning is superior to pointed observations due to its observational efficiency and the high data quality. Drift scanning has already been used in various surveys (*e.g.*, normal galaxies (Shectman *et al.* 1992), low surface brightness galaxies (Dalcanton 1995), and high redshift QSO's (Schneider, Schmidt, and Gunn 1994)).

Standard drift scanning, as described above, is inherently flawed for two reasons: (1) the apparent paths of stars are perfectly straight, and so lie directly along CCD columns, *only* when scanning on the celestial equator, and (2) objects at different declinations, even when the declinations vary by as little as one CCD field-of-view, have noticeably different linear drift velocities. The first difficulty results in images that are at least slightly blurred along CCD rows in any non-equatorial scan. The second difficulty results in images that are at least slightly blurred along CCD columns, because the readout rate is at best only perfectly matched to objects on the central column. The image distortion due to these effects becomes more significant as detectors get larger and as the scan declination deviates from the celestial equator (cf. Gibson & Hickson 1992). Across the fields-of-view covered by today's standard $2048^2$ CCD's with $\sim 1$ arcsec pixels, the image smearing is $\sim 4$ arcsec at a declination of only $\pm 30°$. To avoid blurring the stellar images over more than two pixels, the observations with this system must be restricted to declinations $< 30°$. One solution to this problem is to only use small detectors, vary the readout rate among detectors, and construct the detectors with a curved geometry. The problem with this approach is that the final device is optimized for a single declination.

In this paper we discuss a new instrument, the Great Circle Camera (GCC), that minimizes image distortion by continually reorienting the CCD in such a manner that the scan is conducted along a great circle on the sky rather than on a polar circle of constant declination. In §2 we discuss why great circles are the optimum scan paths and how to calculate the appropriate CCD orientation. In §3 we discuss the design of the GCC. In §4 we discuss how the camera is used, in particular briefly mentioning the interaction between the camera and the computer that controls both the CCD and the GCC, and presenting results from the camera's commissioning run. This instrument creates opportunities for new observations and several new surveys are already underway, including one for low surface brightness galaxies and one of the Magellanic Clouds.

## 2. SCANNING ALONG A GREAT CIRCLE

What is the optimum scan path on the sky? To answer this question, we visualize the situation by considering a fixed sky (*i.e.*, the Earth is not rotating) and a scan that is conducted by moving the telescope. To avoid having celestial objects cross the field at different rates, which would create distorted images, we require that the path not introduce any field rotation. Such paths are only those that follow great circles on the sky (like the celestial equatorial — which is why standard scans along the celestial equator do not suffer from detectable image distortion). Whether the sky or the telescope is moving is only a matter of reference, so the optimum path for real drift scans are also great circles on the sky. The rotation of the Earth does not alter this conclusion, it merely complicates the mechanics of the scan.

A detector that is confined to the focal plane has three degrees of freedom: a motion along one direction, which we shall define to be $X$ motion, a motion along the direction perpendicular to the $X$ motion, which we shall define to be $Y$ motion, and rotation, $\Theta$, about the $Z$ axis (which is perpendicular to both the $X$ and $Y$ axes). We need to track the projection of any path on the sky onto the focal plane by moving the CCD in a way that combines these three motions. We set up the coordinate systems so that at the center of the field the $X$ axis is tangent to the projection of a line of declination (DEC) and that the $Y$ axis is tangent to the projection of a line of right ascension (RA). Because the RA−DEC coordinate system does not project onto a rectilinear coordinate system at the focal plane, the two coordinate descriptions, RA−DEC and $X-Y$, are only identical at the center of the field. During a scan, an object will move in RA, which is principally motion along the $X$ direction, because of the Earth's rotation. However, since RA and $X$ are not the same, as the object moves away from the center of the field it will diverge from the line of constant $Y$ on which it was moving at the field center. Because a CCD column is a line of constant $Y$ and because we need to keep a celestial object on the same CCD column to avoid smearing, we must adjust the CCD's $Y$ position and $\Theta$ orientation to keep an object moving along along constant $Y$. Therefore, we require a stage that will translate the CCD along the $Y$ axis and rotate it about $Z$ during drift scanning. The design and construction of this stage is described in §3. We now discuss the simple mathematics that determine the translation and rotation necessary to follow a great circle at any time, $t$, along the scan.

The basis for this discussion is the spherical right triangle shown in Figure 1. Through any point on the sky there are an infinite number of intersecting great circles. We choose to scan along the great circle that is tangent to the specified declination at the midpoint of the scan (this maximizes the similarity between a standard scan and our GCC scans and minimizes the amount of $X$ motion during a scan). The symbols in Figure 1 are defined as follows: $\phi$ is the declination of the telescope at the midpoint of the scan, T is the point at which the polar circle of declination $\phi$ and our chosen great circle are tangent, $\delta$ is the declination of the CCD at time $t$, $\alpha$ is the angle corresponding to time $t$, $\beta$ is the complement of the rotation angle of the CCD at time $t$, and $a$ is the arc-length along the great circle at time $t$. The known quantities in this problem are $\phi$ and $\alpha$. The declination of the CCD ($\delta$), the arc-length along the great circle ($a$), and the rotation angle of the CCD, $r$ ($\equiv 90° - \beta$), can be evaluated using trigonometrical identities (cf. Smart 1956) and simple algebra:

$$\tan \delta = \tan \phi \cos \alpha \quad (1)$$

$$\tan a = \cos \phi \tan \alpha \quad (2)$$

and

$$\sin r = \sin \phi \sin \alpha. \quad (3)$$

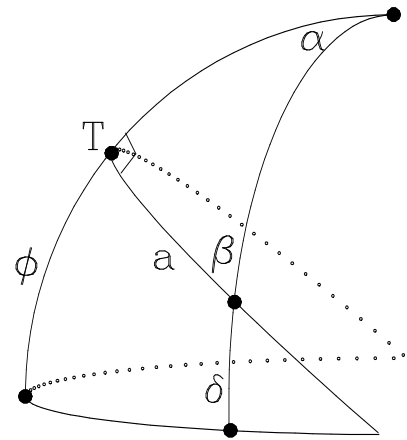

Figure 1. Spherical right triangle. The labels are discussed in the text.

By solving these three equations, the position and rotation of the CCD at any time during a scan is fully specified.

## 3. DESIGN

The principal design drivers are rigidity and compactness. The dewar must be rotated and translated to positional tolerances that are significantly less than the size of a single pixel, which is typically about $20\mu m$. Mechanical advantage on the rotational precision is gained by moving the rotation mechanism to large radius. No such advantage can be gained for the translation motion, and so the translation mechanics require greater precision. The last general consideration is that the shutter and filter assembly must rotate with the CCD so that correspondence is maintained between the CCD and filter. We further require that the camera be able to complete at least a 15 minute scan at the declination of the Magellanic Clouds ($\delta \sim -70°$). Mechanical clearances were set to allow at least 5 cm travel in $X$ and $15°$ in $\Theta$. The achieved rotation and translation clearances in fact allow a significantly longer scan (we discuss a 25 min scan of the Large Magellanic Cloud (LMC) below). There are two limits to the angle through which one can scan. First, there are hard limits imposed by the mechanical design. For example, our rotation mechanism utilizes a tangential arm (see below for more details) that is unable to drive the instrument through large angles and there are considerations regarding clearances. Second, driving the instrument through large angles also implies through large displacements. The CCD must remain in an area where there are only small distortions in the focal plane for scanning to be effective. Therefore, any observing will involve a periodic reset of the instrument to avoid large angle rotations and displacements.

The final basic design is shown in Figure 2 and represents our incorporation of these design goals. The camera assembly is bolted to the back of the telescope through the Telescope Mounting Plate (Plate A). This plate is fixed and has crossed roller slides attached at four points. The slides are mounted as evenly around and as close to the rotation axis as possible to minimize flexure. The slides bear the weight of the instrument and so must be sufficiently strong to carry the instrument without significant deformation. There is a design compromise between the lengths of the four slides (their strength depends on their length) and the rotational clearance necessary for the Filter/Shutter slide (Plate B). The length of the four slides vary between 5 and 12.7 cm in length and are attached to the Translation Plate (Plate C) in such a pattern to provide rotational clearance for Plate B.

The four slides that were used easily support the weight of the instrument below the mounting plate, which is less than 50 kg. Additionally, the slides act as spacers to provide the necessary vertical clearance (1.5 cm) for the 1.3-cm thick Plate B. Plate A is 1.9-cm (0.75 inches) thick and Plates C and D are are 1.3-cm (0.5 inches) thick.

Plate C is attached to Plate A by the four crossed-roller slides and by a lead-screw nut assembly that is fixed onto Plate A. Because Plate A is fixed onto the telescope and onto the lead-screw nut assembly, driving the lead screw translates the portion of the instrument below Plate A relative to Plate A and the telescope. The screw is a stainless-steel 3/8 inch (0.95 cm) outer diameter screw with 16 threads per inch. One end of the screw is fixed and the other end is mounted on a movable plate so that fine adjustments on the parallelism between the screw and Plate A can be made on assembly. The screw is driven through a 2:1 chain-sprocket combination by a 5-phase stepper motor with a maximum holding torque of 51.4 oz-in and 500 steps/revolution. The 5-phase motors were chosen over the conventional 2-phase motors because shaking induced by conventional stepper motors was a concern. The use of the 5-phase motors reduces the torque ripple from 29% of the maximum torque to 5% and we have observed no adverse effects from motor-induced shaking.

Plate B has v-guide tracks (not shown in Figure 2) on both edges (left and right edges of Plate B as seen in Figure 2) and is supported by four v-guide wheels (also not shown in Figure 2, although their support posts fit through the four "semi-circular" holes shown in Plate C that surround the large central opening). The bushings for the wheels have off-center mounting holes that allow one to adjust the final fit between the v-wheels and the track by rotating the bushing. These wheels, via their supporting posts, are mounted onto the Rotational Plate (Plate D). Clearance holes in Plate C allow these posts to rotate along with Plate D and the CCD.

A gear rack (not shown in Figure 2) is located on the left side of Plate B as drawn in Figure 2, and lies above the v-track so that Plate B can be driven to seven distinct allowed positions. The Plate's allowed positions include the two filter positions (where either filter is directly above the CCD), three closed shutter positions (where the plate occults the CCD), and the two outermost positions along the Plate from which one can access one or the other filter. Because the motor assembly cannot be placed directly below the position of the rack (due to the presence of the CCD dewar), Plate B is driven by a gear that is linked by a chain to an offset drive shaft. This chain and the associated sprockets fit within the vertical clearance created by the height of the crossed-roller slides. The drive shaft, whose other end is secured in the filter/shutter assembly box, also holds a detent wheel. Although the ideal situation is to have the motor, the detent, and the gear that drives the rack all along one shaft in order to maximize the precision of the slide's position, we have not experienced problems with this arrangement. The drive shaft is driven through a 4.5:1 gear ratio by the same type of stepper motor that is used in the translational motion. The filter/shutter assembly box also contains an idler shaft that holds coded gear blanks (the code consists of slits cut into wheels) that are read by slotted optical switches mounted on the inside walls of the box. At least one of the three switches reads an open slit position at each allowed position of Plate B. The five "observing" positions (closed shutter, filter 1, closed, filter 2, closed) are identified by unique combinations of codes. The stepper motor is sufficiently reliable (skipped steps are rare) that we can move accurately from one allowed position to the next by specifying the number of motor steps. The readings from the optical switches are then used to confirm the location of Plate B. The motor windings are then turned off and the detent holds Plate B during the exposure and ensures positional repeatability.

Figure 2. Exploded schematic view of the Great Circle Camera with the telescope at the top of the Figure (not shown) and the CCD dewar at the bottom. The drawing is nearly to scale. The entire assembly including Plate A, but not the CCD dewar, is between 6.5 and 7.5 cm (2.5 and 3 inches) thick. Arrows placed by Plates B, C, and D indicate the direction of possible motion for the relevant Plate.

Plate B, which houses two filters (up to 5-inch round filters) takes about 2 sec to move from one allowed position to the next. Because a scan exposure officially begins after the first packet of electrons have crossed the entire field-of-view and stops when the last line is read out, fast and precise shutter timing is not necessary. For each exposure, Plate B will move to the open position by moving the correct filter in from one side and to the closed position by moving the filter out the opposite side. This pattern assures roughly equal exposure time across the CCD for snap (pointed) observations. However, one cannot use bright standards (exposure times $\lesssim$ few seconds) nor can one rely on the shutter to provide exposure times to 1% for exposures that are shorter than 200 seconds. Therefore, standards must be taken in a two-step process where one calibrates the exposure time of short exposures using faint field stars that are also observed in scan mode, for which the exposure time is precisely known to be the time it takes an object at that declination to cross the field-of-view.

Also attached to Plate D is the rotation assembly box, which contains a lead-screw mechanism that drives the rotation of Plate B and the instrument below Plate C relative to the telescope. Plate C and D are connected with a circular bearing of the size of the large central opening (22.9 cm or 9 inches diameter) shown in Figure 2. The linear lead-screw motion is converted into a rotational motion through the use of a flexible connection assembly between the lead-screw nut and Plate C. This flexible connection consists of two aluminum plates that are about two by two inches in size, are within the rotation assembly box, and are oriented parallel to each other only at the position defined to be zero angle rotation. (hereafter these plates are referred to as "miniplates" to clarify the distinction between them and the large plates shown in Figure 2). One of the miniplates is affixed to Plate C, and the other to the lead-screw nut. The one miniplate that is attached to Plate C is shown in Figure 2 to be protruding from a clearance hole at the top of the rotation assembly box. The two miniplates are connected to each other with a set of three thin spring steels strips. The three strips are arranged so that their long axis is parallel to the lead-screw. For each strip, one end is mounted onto one of the two miniplates and the other end onto the other miniplate. This allows the two miniplates to link the motion from the rotation assembly box to Plate C, despite the fact that the miniplates are separating and changing orientation with respect to each other as the instrument rotates. Because the lead screw is placed 25.4 cm (10 inches) away from the rotation axis and the CCD edge is only slightly over 2.5 cm away from the same axis, the rotational precision only needs to be accurate to a few tens of microns. The lead screw, which has the same properties as that used for translation, is again fixed on one end and allowed slight movement during assembly at the other end in order to assure parallelism. The screw is driven through a 1:1 chain/sprocket combination by a 5-phase stepper motor.

The position of the stage in $X$ and $\Theta$ are measured using two different precision positional gauges. The translation gauge has a resolution of $2\mu$m and 50 mm travel. Its body is mounted onto Plate A. The gauge's probe is unattached to Plate B but has a measuring force of between 2 and 4 oz, which keeps the edge of the probe resting on the edge of Plate B. The rotation gauge has a resolution of 20 $\mu$m and 100 mm travel. Its body is attached to Plate D and its tip is attached to the lead-screw nut. The nut's linear position along the screw is translated into the corresponding CCD rotation angle by using simple trigonometric identities and by setting the radius of the circle traced by the rotation to be the distance between the CCD center and the position of the outer side of the innermost miniplate. Again, because of the advantage gained by placing the lead-screw at large radius, lower precision is required in describing this motion.

A single PC is used to manage the camera motion and CCD. The gauge output signals are converted to RS232 and then directly read by the PC through the RS232 input. Each stepper motor (a total of three) has a driver that interprets pulse signals and has opto-isolated input and output. The motor drivers and the three optical switches that specify the position of Plate B are controlled through a single PC parallel port. The instrument and the box that contains the drivers are mounted on the side of the telescope.

4. USING THE GCC

The key to the successful operation of the GCC is the reference CCD position. Initial setup requires iterative pointing and rotation adjustments so that the central CCD column and a line of constant declination are tangent at the central CCD row (for scanning, columns must lie along the E-W axis). Alternatively, this alignment can be stated as requiring that the center of the CCD be the one point for which the $X-Y$ and RA−DEC coordinate systems are identical.

An empirical determination of the sky drift rate is necessary because the physical pixel size and the absolute calibration of the computer's internal clock are not precisely known. The optimum readout rate is determined by minimizing the image distortion along columns (RA). This is a trial-and-error process. Several iterations of the adjustment of both rotation and readout rate are necessary. The scan rate will scale inversely with $\cos(\delta)$ for small differences in $\delta$.

The gross instrument rotation is adjusted with the telescope mounting ring. Finer adjustments are done by redefining the zero point of the rotation gauge. Fine tuning is done by finding a rotation angle that minimizes the image distortion along rows (DEC) for objects in a low-declination scan. This is also a trial-and-error process that is best done once the drift rate is fairly well-determined.

For the best image quality, the motion of the camera must be precisely linked to the CCD readout. To ensure synchron-icity, a single computer and a single associated internal clock are used to time both the readout rate and the camera motion. To avoid computation during the scan, a look-up table of positions and rotation angles as a function of time during the scan is calculated before the scan begins. As each line is read, the computer must decide whether a move is required. If the position of the stage as measured with the gauges differs significantly from the position in the lookup table, then a move is necessary. The move is executed by sending step commands to the stepper motors. The measured position of the stage is compared to the predicted position. Fine tuning of the position is executed by a three-step iterative closed-loop.

The only information required to begin a scan is the length of the scan and the midpoint declination of the scan (hundreths of a degree is sufficient precision). The PC then drives the camera to the initial position (maximum rotation and most northernly declination — for a southern hemisphere scan). The camera will move south during the exposure and then eventually back north to the same declination as the initial position, but it will be oriented with the opposite rotation angle. Once the camera is at its initial position, the shutter opens, the CCD begins scanning, and the stage begins moving. The exposure officially begins (data are saved) once a row of charge has traversed the entire CCD.

The GCC camera had its commissioning run in September 1994 at the Las Campanas 1-m telescope. To test the camera we conducted several 2° scans of the Clouds, each of which lasted about 25 minutes. The central coordinates of the westerly edge of the scan shown in Figure 3 are $5^h\ 27^m\ 35.4^s$ and $-69°\ 01'\ 28''$ (Epoch 1994). This position was chosen so that the scan would include the 30 Doradus region. The scan shown in Figure 3 is 10285 lines long and the readout time was 0.130450 seconds per line (a total exposure time of 22.4 minutes and an effective exposure time on any particular object of about 4.5 minutes). For comparison, we also conducted a standard scan (no translation or rotation) of part of the same region. A comparison of a small portion of the scan shown in Figure 3 to the standard scan of the same region is shown in Figure 4. The atmospheric conditions were worse for the standard scan, so one should compare only the image quality and not the limiting magnitudes of the two panels. Nevertheless, the improvement achieved by scanning along a great circle is striking. The image from the standard scan has gross distortions that render it nearly scientifically useless.

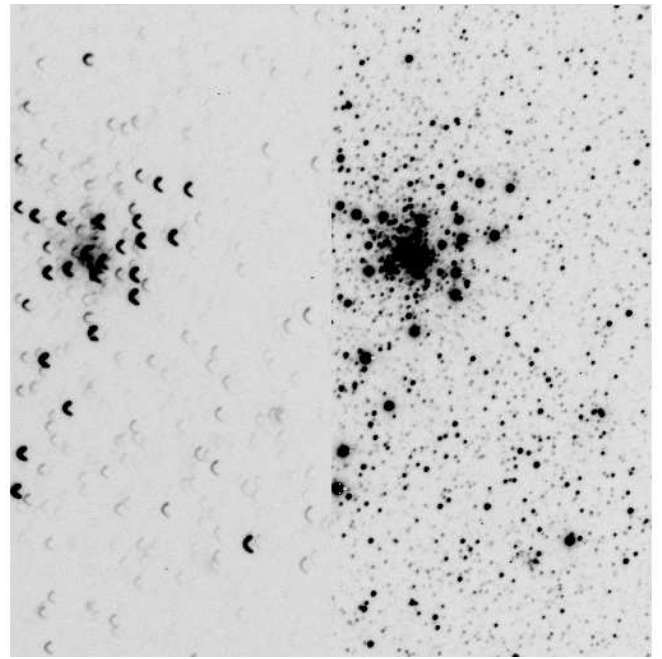

Figure 3. A 2° by 24 arcmin scan of the Large Magellanic Cloud including the 30 Doradus region taken with the GCC. North is to the left and East is up.

Figure 4. A portion of the 2° LMC scan taken with the GCC is shown in the right panel and an uncorrected scan of the same region is shown on the left. The vertical dimension of the image is nearly 6 arcmin.

## 5. SUMMARY

In this paper we discuss the concept of CCD drift scanning along a great circle, explain why this is preferable to standard scanning along circles of constant declination, and discuss the design and construction of the first instrument designed specifically to accomplish this. The Great Circle Camera consists principally of a rotation and translation stage that mounts between the telescope and the CCD. By precisely rotating and translating the CCD, the CCD will track a great circle on the sky for a limited time. The angular length of the allowed scan depends on the declination of the scan, the mechanical design of the camera, and the storage capacity of the observing computer. We present the results of a 2° scan at a declination of $-70°$. A comparison between a standard scan and the new GCC scan demonstrates that the gross distortions of a standard scan at such a declination have been removed through the use of great circle scanning.

This innovation opens new avenues for drift scanning with current telescopes, in particular those with large and flat fields-of-view. The most modern altitude-azimuth telescopes should be able to scan along great circles by utilizing their precise pointing and instrument rotators; however, such a scheme has not yet been implemented. The GCC provides the same capability to existing telescopes at an extremely modest cost with mechanical ease.

Acknowledgments: The material costs of this camera were generously funded by the Dudley Observatory through a Fullam award to D.Z. D.Z. also acknowledges partial financial support provided by NASA through grant HF-1027.01-91A from STScI, which is operated by AURA, Inc., under NASA contract NAS5-26555. The authors also want to thank Oscar Duhalde at Las Campanas for his excellent assistance without which we would not have gotten the instrument on the telescope and the machinists in the Carnegie shops, Pilar Ramirez, Lalo Vasquez, and Joe Dizon for their work in constructing the GCC.